%% file: main.tex
\documentclass[runningheads]{llncs}

\usepackage{cite}
\usepackage{lipsum}
\usepackage{tikz}
\usetikzlibrary{shapes,arrows}
\usepackage{standalone}
\usepackage{adjustbox}
\usepackage{amsmath,amssymb,amsfonts, mathtools}  
\usepackage{algorithmic}
\usepackage{graphicx}
\usepackage{textcomp}
\usepackage{xcolor}
\usepackage{tabularx,booktabs}
\usepackage[font=small,labelfont=bf]{caption} % Required for specifying captions to tables and figures
\usepackage[subrefformat=parens]{subcaption}
\usepackage{hyperref}

\usepackage{pdfpages}
\newcolumntype{C}{>{\centering\arraybackslash}X}
\usepackage{multirow}
\definecolor{color_blue_light}{RGB}{166,206,227}
\definecolor{color_blue}{RGB}{31,120,180}
\definecolor{color_green_light}{RGB}{178,223,138}
\definecolor{color_green}{RGB}{51,160,44}
\definecolor{color_red_light}{RGB}{251,154,153}
\definecolor{color_red}{RGB}{227,26,28}
\definecolor{color_yellow}{RGB}{253,191,111}

\definecolor{color_red_red}{RGB}{228,26,28}
\definecolor{color_blue_blue}{RGB}{55,126,184}
\definecolor{color_green_green}{RGB}{77,175,74}
\definecolor{color_purple_purple}{RGB}{152,78,163}

\definecolor{color_green_blue}{RGB}{102,194,165}
\definecolor{color_orange}{RGB}{252,141,98}
\definecolor{color_gray}{RGB}{141,160,203}

\definecolor{google_green}{RGB}{50,185,100}
\definecolor{google_blue}{RGB}{0,87,231}
\definecolor{google_red}{RGB}{214,45,32}
\definecolor{google_yellow}{RGB}{255,167,0}

\definecolor{color1}{HTML}{ffb85a}
\definecolor{color2}{HTML}{ff655a}
\definecolor{color3}{HTML}{ff5ae9}
\definecolor{color4}{HTML}{705aff}

\begin{document}

\title{Prediction-Coherent LSTM-based Recurrent Neural Network for Safer Glucose Predictions in Diabetic People\\}

\titlerunning{Prediction-Coherent LSTM RNN for Safer Glucose Predictions in Diabetes}

% \href{https://orcid.org/0000-0002-4181-2422}{\includegraphics[scale=0.0045]{figures/orcid.png}}

\author{Maxime De Bois\inst{1}\orcidID{0000-0002-4181-2422} \and Moun{\^i}m A. El Yacoubi \inst{2}\orcidID{0000-0002-7383-0588} \and Mehdi Ammi\inst{3}\orcidID{0000-0003-1763-4045}}

\authorrunning{M. De Bois et al.}

% Mehdi Ammi \href{https://orcid.org/0000-0003-1763-4045}{\includegraphics[scale=0.0045]{graphs/orcid.png}}, and Moun{\^i}m A. El Yacoubi \href{https://orcid.org/0000-0002-7383-0588}{\includegraphics[scale=0.0045]{graphs/orcid.png}}


\institute{CNRS-LIMSI and Universit{\'e} Paris Saclay, Orsay, France \\\email{maxime.debois@limsi.fr} \and Samovar, CNRS, T{\'e}l{\'e}com SudParis, Institut Polytechnique de Paris, \'{E}vry, France \email{mounim.el\_yacoubi@telecom-sudparis.eu}  \and Universit{\'e} Paris 8, Saint-Denis, France \\\email{ammi@ai.univ-paris8.fr}}

% \institue{test \and

% \author{Blind for review.}
% \authorrunning{Blind for review.}
% \institute{Blind for review.}

\maketitle 

\input{0-Abstract.tex}

\input{1-Introduction.tex}

\input{3-pcRNN.tex}

\input{4-Methods.tex}

\input{5-ResultsDiscussion.tex}

\input{6-Conclusion.tex}

\end{document}

%% file: 0-Abstract.tex
\begin{abstract}
In the context of time-series forecasting, we propose a LSTM-based recurrent neural network architecture and loss function that enhance the stability of the predictions. In particular, the loss function penalizes the model, not only on the prediction error (mean-squared error), but also on the predicted variation error.

We apply this idea to the prediction of future glucose values in diabetes, which is a delicate task as unstable predictions can leave the patient in doubt and make him/her take the wrong action, threatening his/her life. The study is conducted on type 1 and type 2 diabetic people, with a focus on predictions made 30-minutes ahead of time.

First, we confirm the superiority, in the context of glucose prediction, of the LSTM model by comparing it to other state-of-the-art models (Extreme Learning Machine, Gaussian Process regressor, Support Vector Regressor).

Then, we show the importance of making stable predictions by smoothing the predictions made by the models, resulting in an overall improvement of the clinical acceptability of the models at the cost in a slight loss in prediction accuracy.

Finally, we show that the proposed approach, outperforms all baseline results. More precisely, it trades a loss of 4.3\% in the prediction accuracy for an improvement of the clinical acceptability of 27.1\%. When compared to the moving average post-processing method, we show that the trade-off is more efficient with our approach.

\end{abstract}

%% file: 1-Introduction.tex
\section{Introduction}

With 1.5 milion inputed deaths in 2012, diabetes is one of the leading diseases in the modern world \cite{world2016global}. Diabetic people, due to the non-production of insulin (type 1) or an increased resistance to its action (type 2), have a lot of trouble managing their blood glucose. In one hand, when their glycemia falls too low (state of hypoglycemia), they are at risk of short-term complications (e.g., coma, death). In the other hand, if their glycemia is too high (hyperglycemia), the complications are long-term (e.g., cardiovascular diseases, blindness).

A lot of efforts are focused towards helping diabetic people in their daily life, with, for instance, continuous glucose monitoring (CGM) devices (e.g., FreeStyle Libre \cite{olafsdottir2017clinical}), artificial pancreas (e.g., MiniMed 670G \cite{messer2018optimizing}), or coaching smartphone applications for diabetes (e.g., mySugr \cite{rose2013evaluating}). Thanks to the advances in the field of machine learning and the increased availability of data, a lot of researchers are following the lead of the prediction of future glucose values. The goal is to build data-driven models that, using the patient's past information (e.g.,  glucose values, carbohydrate intakes, insulin boluses), predict glucose values multiple minutes ahead of time (we call those models multi-step predictive models).

While a lot of the early work in the glucose prediction field were focused on the use of autoregressive (AR) models \cite{sparacino2007glucose}, the models that are used nowadays are more complex. Georga \textit{et al.} explored the use of Support Vector Regression (SVR) in predicting glucose up to 120 minutes ahead of time in type 1 diabetes \cite{georga2013multivariate}. Valletta \textit{et al.} proposed the use of Gaussian Process regressor (GP) to include a measure of the physical activity of type 1 diabetic patients into the predictive models \cite{valletta2009gaussian}. In their work, Daskalaki \textit{et al.} demonstrated the superiority of feed-forward neural networks compared to AR models \cite{daskalaki2012real}. As for them, Georga \textit{et al.} studied the use of Extreme Learning Machine models (ELM) in short-term (PH of 30 minutes) type 1 diabetes glucose prediction \cite{georga2015online}. Finally, recurrent neural networks (RNN) have shown a lot of interest in the field \cite{zarkogianni2011insulin}, and in particular those with long short-term memory (LSTM) units \cite{mirshekarian2017using,sun2018predicting,martinsson2018automatic,debois2018study}.

 However, neural-network-based models, while exhibiting very promising results, often show instability in the predictions. This comes from the training of the models that, most of the time, aims at minimizing the mean-squared error (MSE) loss function. It makes the model focus on getting a good point-accuracy, without questioning the coherence of consecutive predictions.

% \textcolor{cyan}{Towards this goal, the autoregressive model and its derivatives have been widely used in the past \cite{sparacino2007glucose,reifman2007predictive,eren2012adaptive}. However, due to their inherent simplicity, they have recently fallen out of favor either for more complex regression models (e.g., support vector regression \cite{bunescu2013blood,georga2013multivariate}, Gaussian process regression \cite{tomczak2016gaussian,de2015controlling}) or neural-network-based models (e.g., feed-forward neural network \cite{perez2010artificial,daskalaki2012real,zecchin2012neural}, recurrent neural network \cite{daskalaki2013early,mirshekarian2017using}, extreme learning machine \cite{georga2015online}, convolutional neural network \cite{li2018convolutional}).}

The stability of the predictions is very important in predicting future glucose values. Predicting towards the wrong direction or with consecutive inconsistent directions can make the diabetic patient take the wrong action, potentially threatening his/her life. This is why the accuracy of the predicted glucose variations is taken into account when assessing the clinical acceptability of glucose predictive models, with, for instance, the widely-used Continuous Glucose-Error Grid Analysis (CG-EGA) \cite{oviedo2017review}. We identified that this issue is not specific to the field of glucose prediction and can be extended to other multi-step forecasting applications, such as stock market prediction \cite{dong2013one} or flood levels forecasting \cite{chang2007multi}.

% wind speed forecasting \cite{wang2017research}, 
% or the long-term prediction of temperature \cite{duhoux2001improved}

% https://www.sciencedirect.com/science/article/pii/S036054421730333X
% https://ieeexplore.ieee.org/stamp/stamp.jsp?arnumber=6782784
% Multi-step-ahead neural networks for flood forecasting - chang
% Improved  long-term  temperature  prediction  by  chaining  of  neural  networks - duhoux

%Mehdi : ça serait interessant de montrer ce qui t'a ammené à cette approche ? resultats dans d'autres domaines ? demonstration ?
%Mehdi : parler de contribition. Our contribution ...

In this paper, to enhance the stability of the predictions, we propose a new LSTM-based RNN architecture and loss function. We demonstrate the usefulness of the idea by applying it to the challenging task of predicting future glucose values of diabetic patients which directly benefits from an increased stability.

We can summarize our contributions as follows:
\begin{enumerate}
    \item We propose a new loss function that penalizes the model simultaneously during its training, not only on the classical MSE, but also on the predicted variation error. To be able to compute the penalty, we propose to use the loss function in a two-output LSTM-based RNN architecture. We validate the proposed approach by comparing it to four other state-of-the-art models.
    \item We demonstrate the importance of making stable predictions in the context of glucose predictions as accurate but unstable predictions lead the models to have a bad clinical acceptability.
    \item We confirm the overall usefullness of using LSTM-based RNN in predicting future glucose values by comparing it to other state-of-the-art models. In particular, the LSTM model shows more clinical acceptable results.
    \item We have conducted the study on two different datasets, one with type 1 and one with type 2 diabetic patients. This is worth mentioning as glucose prediction studies are very rarely done on type 2 diabetes (although it represents around 90\% of the whole diabetic population).
    \item Finally, we have made all the source code and a standalone implementation of the CG-EGA available in Github.
\end{enumerate}

% two-outputs RNN with a penalty based on the predicted variation error to the traditional mean square error loss function. We apply the idea to the case of multi-step ahead prediction of glucose in diabetes. 

The rest of the paper is organized as follows. First, we introduce the proposed architecture and loss function. Then, we present its application to the prediction of future glucose values. Finally, we provide the reader with the results and takeaways from the experiments.

%% file: 3-pcRNN.tex
\section{Prediction-Coherent LSTM-based Recurrent Neural Network} \label{sec:pcRNN}
%Mehdi : il manque une petite transition entre review et contribution. peut être une courte intro de la cotribution.

\subsection{Presentation of the Model}

In multi-step time-series forecasting, at time $t$, the model takes a set of features $\boldsymbol{X}$ to predict the future value of the time-series $\boldsymbol{y}$ at a prediction horizon $PH$: $\hat{y}_{t+PH}$. Most of the time, the input features $\boldsymbol{X}$ comprises the past $H$ known values of the time-series $\boldsymbol{y}$ as well as other time-related features.

% \textcolor{cyan}{problem of models single output in general}

% In this paper, we call two consecutive predictions, $\hat{y}_{t+PH-1}$ and $\hat{y}_{t+PH}$, coherent with the true values when the variation from one to the other, $\Delta\hat{y}_{t+PH}$, reflect the true variation of the time-series $\Delta y_{t+PH}$.

% Mehdi : c'est cet element qu'il faut évoqué brièvement en intro peut être ? pour la justitification de ton choix

RNN, and in particular those based on LSTM cells, are neural networks that are particularly suited for time-series forecasting as they include the temporal component of the features and the predictions into their architecture \cite{mandic2001recurrent}. Such models are usually trained with the MSE loss function (see Equation \ref{eqn:mse}) which estimates the mean accuracy of the predictions. 

\begin{equation}
    MSE(\boldsymbol{y},\boldsymbol{\hat{{y}}})=\frac{1}{n} \sum_{i=1}^n ({y}_{i}-\hat{{y}}_{i})^2
\label{eqn:mse}    
\end{equation}

However, using the MSE does not incentivize the model to make successive predictions that are coherent with their respective true values. More formally, we can call two consecutive predictions, $\hat{y}_{t+PH-1}$ and $\hat{y}_{t+PH}$, coherent with the true values when the predicted variation from one to the other, $\Delta\hat{y}_{t+PH}$, reflect the true variation of the time-series $\Delta y_{t+PH}$.

%Mehdi : reformuler la partie "a simple idea..." pour mieux valoriser ta contribution. 
To enhance to coherence of consecutive predictions, we propose the idea of using a two-output LSTM that takes advantage of its architecture to penalize incoherent successive predictions during its training. We call this neural network a Prediction-Coherent LSTM-based recurrent neural network (pcLSTM).

%Mehdi : ne pas utilisé simple... standard RNN

\subsubsection{Two-output LSTM.} 
\begin{figure}
\centering
\includegraphics[width=0.7 \textwidth]{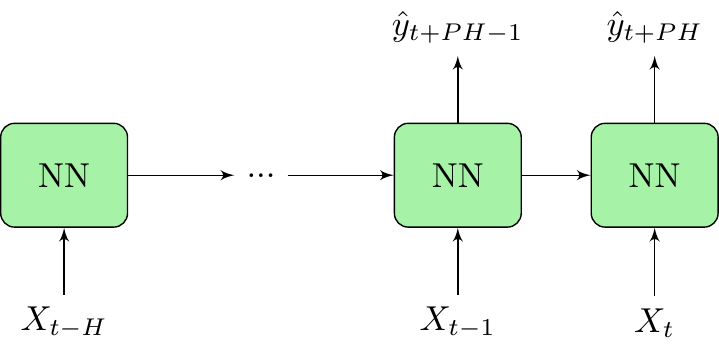}
\caption{two-output LSTM which has been unrolled $H$ times. $X_t$ are the input features at time $t$ and $\hat{y}_{t+PH}$ is the forecast of the time-series $y$ at a time $t+PH$.}
\label{fig:consistent_rnn}
\end{figure}

The two-output LSTM is a standard LSTM unrolled $H$ times and that outputs the predictions of the last two steps (see Figure \ref{fig:consistent_rnn}).

% Because the most recent features are often the most important when making the predictions, the model can display a high sensitivity to incoming features. 

% In particular, when the predictive task is hard (e.g., because of the nature of the time-series of a high PH), successive predictions can display a high sensitivity to the new incoming features.

% However, while predictions of the preceding time-steps are taken into account in the computation of the next prediction, there is no constraint on how

%mehdi : parler de contribution. pour calrifier le message au reviewer. our contribution ... ou un qq chose du meme genre

\subsubsection{Variations Penalized Loss Function.} To enhance the coherence between two consecutive predictions, we propose to penalize the network on the error of the predicted variation.
We define the cMSE (see Equation \ref{eqn:cmse}), which is the weighted sum of the MSE of the predictions and the MSE of the predicted variations. We call the parameter $c$ the \textit{coherence factor}. It represents the relative importance of the variation-based penalty compared to the accuracy of the predictions.

\begin{equation} \label{eqn:cmse}
\begin{split}
cMSE(\boldsymbol{y},\boldsymbol{\hat{y}}) & = MSE(\boldsymbol{y},\boldsymbol{\hat{y}}) + c \cdot MSE(\boldsymbol{\Delta y},\boldsymbol{\Delta \hat{y}}) \\
 & = \frac{1}{n} \sum_{i=1}^n ({y}_i-\hat{{y}}_i)^2 + c \cdot (\Delta{y}_i-\Delta\hat{{y}}_i)^2
\end{split}
\end{equation}

The coherence factor $c$ is a problem-dependent parameter that has to be optimized depending on the relative importance of having coherent or stable predictions versus having accurate predictions.

We note that, if the coherence factor, $c$, is set to $0$, the $cMSE$ becomes the $MSE$ and the model then behaves like a standard one-output LSTM model.

%% file: 4-Methods.tex
\section{Methods}

In this section, we go through the experimental details of the study, and, in particular, the data we used, the preprocessing steps we followed, the models we implemented, and the evaluation metrics we used.

We made the source code used in this study available in the pcLSTM Github repository \cite{debois2019pcLSTM}.

\subsection{Experimental Data}

Our data come from two distinct datasets: the Ohio T1DM dataset and the IDIAB dataset accounting for 6 type 1 and 5 type 2 diabetic patients respectively. 

\subsubsection{Ohio Dataset.} First published for the Blood Glucose Level Prediction Challenge in 2018, the OhioT1DM Dataset comprises data from six type 1 diabetic people who were monitored during 8 weeks \cite{marling2018ohiot1dm}. For the sake of simplicity and the uniformity with the IDIAB dataset, we restrict the dataset to the glucose readings (in $mg/dL$), the daily insulin boluses (in units) and the meal information (in $g$ of CHO).

% The dataset comprises data from six type 1 diabetic people (2M/4F, age between 40 and 60 years old, BMI and HbA1c not disclosed) who were monitored during 8 weeks. During the whole duration, they wore MiniMed\textsuperscript{\textregistered} 530G insulin pumps (Medtronic), Enlite\textsuperscript{\textregistered} CGM sensors (Medtronic). Life-events data have been recorded through a custom smartphone app and physical activity signals have been retrieved from Basis Peak fitness bands (Basis). In this study we only used the CGM glucose readings (in $mg/dL$), the daily insulin boluses (in units) and the meal information (in $g$ of CHO). We chose not to use the other available data (a complete description of the dataset can be found here \cite{marling2018ohiot1dm}) for the sake of simplicity and uniformity with the IDIAB dataset.

\subsubsection{IDIAB Dataset.} For this study, we conducted a data collection on the type 2 diabetic population. The data collection and the use of the data in this study has been approved by the french ethical committee ``Comit\'{e}s de protection des personnes" (ID RCB 2018-A00312-53).

Five people with type 2 diabetes (4F/1M, age 58.8 $\pm$ 8.28 years old, BMI 30.76 $\pm$ 5.14 $kg/m^2$, HbA1c 6.8 $\pm$ 0.71 \%), have been monitored for 31.8 $\pm$ 1.17 days in free-living conditions. The patients were equipped with FreeStyle Libre (FSL) CGM devices (Abbott Diabetes Care) \cite{olafsdottir2017clinical}, which were recording their glucose levels (in $mg/dL$), and with the mySugr (mySugr GmbH) coaching app for diabetes \cite{rose2013evaluating}, in which the patient logged his/her food intakes (in $g$ of CHO) and insulin boluses (in units).

\subsection{Preprocessing.} 

The goal of the preprocessing part is to uniformize the two datasets and prepare them for the training and testing of the models.

\subsubsection{Data Cleaning.} To balance the training and the testing sets regarding the distribution of the samples on the daily timeline, we have chosen to remove incomplete days from the datasets. As a result, for every patient, we ended up with an average of 38.5 ($\pm$ 4.82) and 29.4 ($\pm$ 1.62) days worth of data for the Ohio and IDIAB datasets respectively.

We noticed that several glucose readings in the IDIAB dataset were erroneous (characterized by high amplitude spikes). As this is not particularly surprising (a study by Fokkert \textit{et al.} reported that only 85.5\% of the FSL readings were within $\pm20\%$ of the reference sensor values \cite{fokkert2017performance}), we removed them to prevent them from disturbing the training of the model.

\subsubsection{Resampling \& Interpolation.} To synchronize the data between them, we have resampled both datasets to get a sample every 5 minutes. During the resampling process, glucose values have been averaged, insulin boluses and CHO intakes have been summed up.

To make up for the introduced missing glucose values in the IDIAB dataset (which has one reading every 15 minutes, instead of 5), we interpolated the glucose signals as it has already been done in the context of glucose prediction \cite{stahl2008short}. In particular, we used a piecewise cubic hermite interpolating polynomial (PCHIP) \cite{fritsch1980monotone} to avoid oscillations in the interpolated signal (which occurred with a single polynomial interpolation) and to preserve the monotonicity of the fitted signal (which was an issue with a spline interpolation) \cite{li2007greater}.

\subsubsection{Datasets Splitting.} To ready up the datasets for the training and testing  of the models, we have to create the training, validation and testing sets. The splitting of the data has been done on full days of data to ensure an uniform distribution of the daily sequences across the datasets. We split the data into training, validation and testing sets following a 50\%/25\%/25\% distribution.

\subsubsection{Input Scaling.} Lastly, the training sets data have been standardized (zero-mean and unit-variance). The same transformation has then been applied to the validation and testing sets.

\subsection{Models} In this study, we compare the proposed approach (pcLSTM) to four other state-of-the-art models, namely  an Extreme Learning Machine neural network (ELM), a Gaussian Process regressor (GP), a LSTM recurrent neural network (LSTM), and a Support Vector Regression model (SVR).

Every model is personalized to the patient. To be able to model long-term dependencies, every model takes the past 3 hours of glucose, insulin, and CHO values as input. The hyperparameters of every model have been tuned on the validation sets by grid search.

\subsubsection{ELM.} The ELM architecture has $10^5$ neurons in its single hidden layer. To reduce the impact of overfitting, we applied a L2 penalty ($500$) to the weights.

\subsubsection{GP.} The GP model has been implemented with a dot-product kernel. The dot-product has been chosen instead of a traditional radial basis function kernel as it has been shown to perform better in the context of glucose prediction \cite{debois2018study}. The inhomogeneity parameter of the kernel has been set to $10^{-8}$. To ease the fitting of the model, white noise (value of $10^{-2}$) has been added to the observations.

\subsubsection{LSTM.} The LSTM model is made of a single hidden layer of 128 LSTM units. It has been trained to minimize the MSE loss function using the Adam optimizer with batches of 10 samples and a learning rate of $5\times10^{-3}$. To prevent the overfitting of the network to the training data, we added a L2 penalty ($10^{-4}$) and used the early stopping methodology.

\subsubsection{pcLSTM.} The pcLSTM recurrent neural network shares the same characteristics with the LSTM model. The only difference is its two-output architecture and its associated cMSE loss function (see Section \ref{sec:pcRNN}). In particular, the coherence factor has been optimized through grid search to ensure a good trade-off between the accuracy of the predictions and the accuracy of the predicted variations. We settled down with a coherence factor of $2$.

\subsubsection{SVR.} The SVR model has been implemented with a radial basis function (RBF) kernel. The coefficient of the kernel has been set to $5 \times 10^{-4}$. The wideness of the no-penalty tube has been set to $0.1$ and the penalty itself has been set to $50$.

\subsection{Post-processing} \label{sec:postprocessing}

By using the cMSE loss function, we incentivize the model to make consecutive predictions reflecting the actual glucose rate of change. In a way, it can be viewed as a smoothing effect integrated to the training of the model.

Some post-processing time-series smoothing techniques exist, such as the exponential smoothing or the moving average smoothing \cite{shumway1982approach}. The latter, yielding a better trade-off between the accuracy of the predictions and the accuracy of the predicted variations, has been used with a window of the last 3 predictions.

\subsection{Evaluation Metrics} \label{sec:metrics}

In this study, three evaluation metrics have been used: the Root-Mean-Squared prediction Error (RMSE), the Root-Mean-Squared predicted variation Error (dRMSE), and the Continuous Glucose-Error Grid Analysis (CG-EGA) measuring the clinical acceptability of the predictions. 

%Mehdo : le RMSE est assez cournant non ? peut être le développer moins que le dRMSE et le CGEGA?

\subsubsection{RMSE.} The RMSE is the most used metric in the world of glucose prediction as it measures the overall accuracy of the predictions \cite{oviedo2017review}.

\subsubsection{dRMSE.} We call the dRMSE the RMSE applied to the difference between two consecutive predictions. Therefore, it measures the accuracy of the predicted variations and can be used to estimate the impact of the variation-based penalty in the cMSE loss function.

% \subsubsection{RMSE.} The RMSE (measured in $mg^2/dL^2$) is one of the most popular metrics in measuring the accuracy of the predictions made by regression models. It is widely used in the world of glucose prediction since, compared to other standard metrics (e.g, mean absolute error), it penalizes large errors harder, which is desirable in predicting future glucose values as large errors can threaten the life of the patient \cite{oviedo2017review}.

% \subsubsection{dRMSE.} We call the dRMSE (measured in $mg^2/dL^2/min^2$) the RMSE applied to the variations of consecutive predictions instead of the RMSE applied to the glucose predictions themselves. More precisely, it is defined by Equation \ref{eqn:drmse}, where $D$ is the number of days worth of data in the testing set, $N$ is the number of samples per day, and $\Delta y$ and $\Delta \hat{y}$ are respectively the variations of the true and predicted glucose values. This metric gives us a way to estimate the impact of the variation-based penalty in the cMSE loss function.

% \begin{equation}
%     dRMSE(\boldsymbol{y},\boldsymbol{\hat{{y}}})=\sqrt{\frac{1}{D \times (N-1)} \sum_{d=1}^{D} \sum_{n=1}^{N-1} ({\Delta y}_{d,n}-\Delta\hat{{y}}_{d,n})^2}
% \label{eqn:drmse}    
% \end{equation}

\subsubsection{CG-EGA.} The CG-EGA provides a measure of the clinical acceptability of the predictions \cite{oviedo2017review}. Indeed, predictions, depending on the current state of the patient's glycemia (hypoglycemia, euglycemia\footnote{The euglycemia region is the region between hypoglycemia and hyperglycemia.}, or hyperglycemia), can be more or less dangerous, which is not taken into account in metrics such as the RMSE.

%Mehdi : est-il possible d'evoquer un peu de literaurre pour montrer l'interet de cette mesure et son impact ?

Technically, the CG-EGA is made of two grids: the Point-Error Grid Analysis (P-EGA) and the Rate-Error Grid Analysis (R-EGA). Whereas the P-EGA provides an acceptability score (from A to E) to the glucose predictions, the R-EGA gives each prediction a score (also from A to E) based on the variation from the previous prediction to the current one \cite{kovatchev2004evaluating}. The CG-EGA combines both grids and gives, for every prediction, in its simplified representation, a clinical acceptability category: accurate prediction (AP), benign error (BE), or erroneous prediction (EP). For a prediction to be categorized as an AP, it needs to have a score of A or B in both the P-EGA and the R-EGA.

We published the source code of the CG-EGA implementation in Github \cite{debois2019CGEGA}.

%% file: 5-ResultsDiscussion.tex
\section{Results and Discussion}

\begin{figure}[!ht]
\centering\begin{adjustbox}{width=0.70\linewidth}
\includegraphics[width=\linewidth]{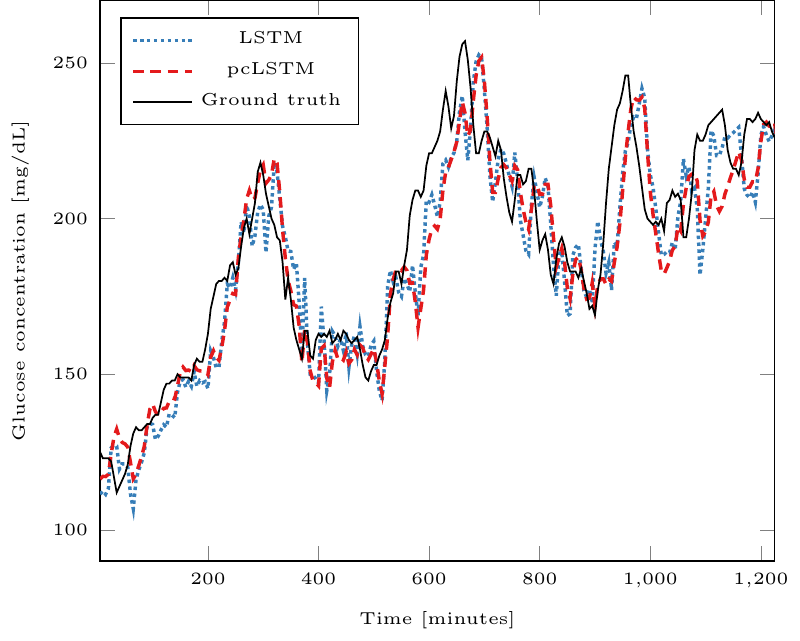}
\end{adjustbox}
\caption{Glucose predictions of the unsmoothed LSTM and pcLSTM against the ground truth, for a given day of one of the patients.}
\label{fig:graph}
\end{figure}

\begin{figure}[!ht]
\centering\begin{adjustbox}{width=0.7\linewidth}
\includegraphics[width=\linewidth]{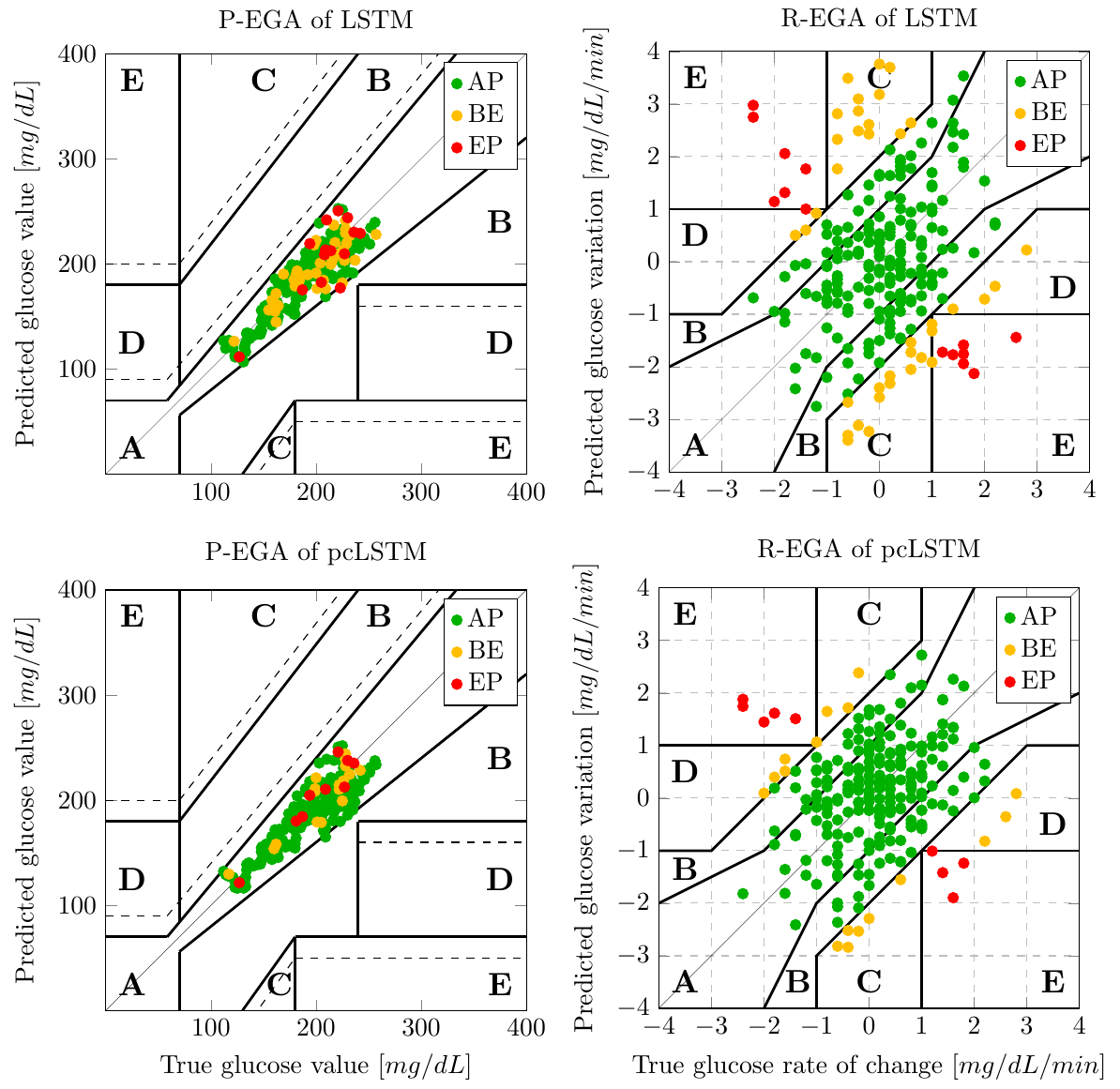}
\end{adjustbox}
\caption{P-EGA (left) and R-EGA (right) for LSTM (top) and pcLSTM (bottom) models for a patient during a given day. The CG-EGA classification (AP, BE, or EP) is computed by combining both P and R-EGA ranks.}
\label{fig:cg_ega}
\end{figure}

The results of the models, presented with and without the moving average smoothing technique discussed in Section \ref{sec:postprocessing}, are reported in Table \ref{table:res_ph30}. Figure \ref{fig:graph} gives a graphical representation of the effect of the proposed approach on the predictions. A detailed graphical clinical acceptability classification of the predictions is given by Figure \ref{fig:cg_ega}.

\input{tables/results_ph-30.tex}

First, when looking at the unsmoothed baseline results, apart from the ELM model that has overall the worse performances (excluding it from the following analysis), we can see that the models have different strengths and weaknesses. Whereas the GP model stands out as being the most point-accurate model (RMSE), it is also the most unstable model (dRMSE). This makes it the least clinically acceptable model of the remaining three, having the lowest AP and the highest EP rates. On the other hand, the SVR model has the worse RMSE, the best dRMSE, and the best AP and EP rates, making it the most clinically acceptable baseline model. Finally, the LSTM model displays competitive results with respect to the GP and SVR models, which validates the use of the LSTM model in the context of glucose prediction.

% In the other hand, we have the LSTM model which makes the least point-accurate predictions (RMSE) but manages to capture the true glucose variations in its predictions the most accurately (dRMSE). At the same time, the LSTM model has the best clinical acceptability with the highest AP and lowest EP rates of the four baseline models. This confirms the importance of having accurate predicted glucose variations and validates the use of the LSTM model in the context of glucose prediction. Finally, the ELM model displays interesting results, which are competitive to the other models (in particular, second best AP rate) while being the easiest model to train and tune.

When looking at the unsmoothed performances of the pcLSTM model, we can see that, compared to the LSTM model, its RMSE is slightly worse ($+4.3\%$), its dRMSE drastically improved ($-24.6\%$) and so is its clinical acceptability ($+27.1\%$ and $-12.8\%$ for the room for improvement in the AP and EP rates respectively). This shows the importance of focusing on the coherence of successive predictions as the increased accuracy in predicted variations (dRMSE) is the main contributor to the increased clinical acceptability.

The results of the models with smoothed predictions show us the general benefit of improving the stability of the predictions to make them more clinically acceptable. Even though all the models see their clinical acceptability improved, the improvement varies from model to model: the models with the highest instability benefit from the smoothing the most. In average, the improvement due to the smoothing applied on the baseline models (still excluding the ELM model) is of $+8.5\%$, $-24.3\%$, $+26.0\%$, and $-14.14\%$ in RMSE, dRMSE, AP and EP rates respectively. Those results show us that the trade-off made by the pcLSTM is much more efficient ($+8.5\%$ against $+4.3\%$ in RMSE for overall the same improvement in the other metrics).

%% file: tables/results_ph-30.tex
% (.*)\t(.*)\t0,(\d\d)(.*)\t0,(\d\d)(.*)\t0,(\d\d)(.*)
% \1 & \2 & \3.\4 & \5.\6 & \7.\8 \\\\

\begin{table*}
    \caption{Performances of the ELM, GP, LSTM, pcLSTM, and SVR models, evaluated at a prediction horizon of 30 minutes with and without the smoothing of the predictions (mean $\pm$ standard deviation, averaged on the subjects from both datasets).}
    \label{table:res_ph30}
    \begin{tabularx}{\linewidth}{C||C|C|C|C|C}
        \toprule
        \multirow{2}{*}{\textbf{Model}} & \multirow{2}{*}{\textbf{RMSE}} & \multirow{2}{*}{\textbf{dRMSE}} &  \multicolumn{3}{c}{\textbf{CG-EGA}} \\
        & & &  AP & BE & EP \\
        \midrule
         
         \multicolumn{6}{c}{\textit{Without smoothing}}\\
         
         \midrule
         
        % \textbf{ELM} & 23.16\scriptsize{ $\pm$ 4.90~} & ~1.70\scriptsize{ $\pm$ 0.43~} & 83.45\scriptsize{ $\pm$ 5.98~} & ~11.56\scriptsize{ $\pm$ 4.17~} & ~4.99\scriptsize{ $\pm$ 2.07~} \\
        
        % \textbf{GP} & \textbf{21.69}\scriptsize{ $\pm$ 4.95~} & ~2.27\scriptsize{ $\pm$ 0.46~} & 80.05\scriptsize{ $\pm$ 5.93~} & ~14.68\scriptsize{ $\pm$ 3.95~} & ~5.27\scriptsize{ $\pm$ 2.22~} \\
        
        % \textbf{LSTM} & 23.85\scriptsize{ $\pm$ 5.07~} & ~1.51\scriptsize{ $\pm$ 0.32~} & 85.70\scriptsize{ $\pm$ 3.77~} & ~9.61\scriptsize{ $\pm$ 2.57~} & ~4.69\scriptsize{ $\pm$ 1.33~} \\
        
        % \textbf{pcLSTM} & 22.95\scriptsize{ $\pm$ 4.84~} & ~\textbf{1.22}\scriptsize{\textbf{ $\pm$ 0.27~}} & \textbf{89.77}\scriptsize{\textbf{ $\pm$ 4.08~}} & ~\textbf{6.55}\scriptsize{\textbf{ $\pm$ 3.03~}} & ~\textbf{3.67}\scriptsize{\textbf{ $\pm$ 1.66~}} \\
        
        % \textbf{SVR} & 22.45\scriptsize{ $\pm$ 4.85~} & ~1.82\scriptsize{ $\pm$ 0.46~} & 82.52\scriptsize{ $\pm$ 5.86~} & ~12.56\scriptsize{ $\pm$ 3.97~} & ~4.92\scriptsize{ $\pm$ 2.03~} \\
        
         \textbf{ELM} & 25.54\scriptsize{ $\pm$ 5.02~} & ~1.90\scriptsize{ $\pm$ 0.45~} & 79.34\scriptsize{ $\pm$ 7.53~} & ~14.92\scriptsize{ $\pm$ 5.50~} & ~5.74\scriptsize{ $\pm$ 2.33~} \\
        
        \textbf{GP} & \textbf{18.92}\scriptsize{\textbf{ $\pm$ 4.56~}} & ~2.21\scriptsize{ $\pm$ 0.44~} & 81.70\scriptsize{ $\pm$ 6.21~} & ~13.88\scriptsize{ $\pm$ 4.09~} & ~4.41\scriptsize{ $\pm$ 2.28~} \\
        
        \textbf{LSTM} & 19.48\scriptsize{ $\pm$ 4.42~} & ~1.95\scriptsize{ $\pm$ 0.40~} & 82.98\scriptsize{ $\pm$ 5.65~} & ~12.42\scriptsize{ $\pm$ 3.73~} & ~4.60\scriptsize{ $\pm$ 2.06~} \\
        
        \textbf{pcLSTM} & 20.32\scriptsize{ $\pm$ 4.56~} & ~\textbf{1.47}\scriptsize{\textbf{ $\pm$ 0.31~}} & \textbf{87.60}\scriptsize{\textbf{ $\pm$ 4.74~}} & ~\textbf{8.76}\scriptsize{\textbf{ $\pm$ 3.23~}} & ~\textbf{4.01}\scriptsize{\textbf{ $\pm$ 1.84~}} \\
        
        \textbf{SVR} & 20.08\scriptsize{ $\pm$ 4.24~} & ~1.74\scriptsize{ $\pm$ 0.44~} & 83.92\scriptsize{ $\pm$ 6.10~} & ~11.75\scriptsize{ $\pm$ 4.24~} & ~4.32\scriptsize{ $\pm$ 2.01~} \\
        
         \midrule
         
         \multicolumn{6}{c}{\textit{With smoothing}}\\
         
         \midrule

        % \textbf{ELM} & 24.75\scriptsize{ $\pm$ 5.06~} & ~1.42\scriptsize{ $\pm$ 0.32~} & 86.72\scriptsize{ $\pm$ 4.87~} & ~8.85\scriptsize{ $\pm$ 3.47~} & ~4.43\scriptsize{ $\pm$ 1.72~} \\
        
        % \textbf{GP} & \textbf{23.10}\scriptsize{\textbf{ $\pm$ 5.10~}} & ~1.54\scriptsize{ $\pm$ 0.32~} & 85.74\scriptsize{ $\pm$ 4.44~} & ~9.86\scriptsize{ $\pm$ 3.10~} & ~4.40\scriptsize{ $\pm$ 1.65~} \\
        
        % \textbf{LSTM} & 25.71\scriptsize{ $\pm$ 5.38~} & ~1.41\scriptsize{ $\pm$ 0.30~} & 86.60\scriptsize{ $\pm$ 4.52~} & ~8.72\scriptsize{ $\pm$ 3.11~} & ~4.68\scriptsize{ $\pm$ 1.83~} \\
        
        % \textbf{pcLSTM} & 25.15\scriptsize{ $\pm$ 5.19~} & \textbf{~1.21}\scriptsize{\textbf{ $\pm$ 0.27~}} & \textbf{89.50}\scriptsize{\textbf{ $\pm$ 4.27~}} & \textbf{~6.62}\scriptsize{\textbf{ $\pm$ 3.08~}} & \textbf{~3.88}\scriptsize{\textbf{ $\pm$ 1.77~}} \\
        
        % \textbf{SVR} & 24.11\scriptsize{ $\pm$ 5.05~} & ~1.48\scriptsize{ $\pm$ 0.33~} & 86.24\scriptsize{ $\pm$ 4.89~} & ~9.44\scriptsize{ $\pm$ 3.43~} & ~4.32\scriptsize{ $\pm$ 1.68~} \\
        
        \textbf{ELM} & 26.64\scriptsize{ $\pm$ 5.17~} & ~1.42\scriptsize{ $\pm$ 0.31~} & 86.13\scriptsize{ $\pm$ 5.26~} & ~9.31\scriptsize{ $\pm$ 3.89~} & ~4.57\scriptsize{ $\pm$ 1.79~} \\
        
        \textbf{GP} & \textbf{20.42}\scriptsize{\textbf{ $\pm$ 4.70~}} & ~1.48\scriptsize{ $\pm$ 0.31~} & 87.17\scriptsize{ $\pm$ 4.38~} & ~9.08\scriptsize{ $\pm$ 3.03~} & \textbf{~3.74}\scriptsize{\textbf{ $\pm$ 1.65~}} \\
        
        \textbf{LSTM} & 21.21\scriptsize{ $\pm$ 4.63~} & ~1.41\scriptsize{ $\pm$ 0.30~} & 87.51\scriptsize{ $\pm$ 4.45~} & ~8.60\scriptsize{ $\pm$ 3.15~} & ~3.89\scriptsize{ $\pm$ 1.60~} \\
        
        \textbf{pcLSTM} & 22.42\scriptsize{ $\pm$ 4.85~} & \textbf{~1.29}\scriptsize{\textbf{ $\pm$ 0.28~}} & \textbf{88.82}\scriptsize{\textbf{ $\pm$ 4.43~}} & \textbf{~7.36}\scriptsize{\textbf{ $\pm$ 3.18~}} & ~3.81\scriptsize{ $\pm$ 1.68~} \\
        
        \textbf{SVR} & 21.81\scriptsize{ $\pm$ 4.43~} & ~1.42\scriptsize{ $\pm$ 0.32~} & 87.38\scriptsize{ $\pm$ 4.78~} & ~8.79\scriptsize{ $\pm$ 3.37~} & ~3.83\scriptsize{ $\pm$ 1.63~} \\
         
        \bottomrule
    \end{tabularx}
\end{table*}

%% file: 6-Conclusion.tex
\section{Conclusion}

In this paper, we have presented a new loss function for recurrent neural networks which, by penalizing the model on the predicted variation errors in addition to the prediction errors, helps the network making more stable predictions.

We apply the proposed model to the prediction of future glucose values in diabetes. First, we validate the use of recurrent neural networks (in particular with LSTM units) by showing that our baseline LSTM model is competitive when compared to other state-of-the-art models. Then, we demonstrate the importance of the proposed approach as it greatly improves the clinical acceptability of the predictions. Lastly, we compare the proposed approach to another smoothing technique. While the effect on the clinical acceptability is the same, the loss in the accuracy of the prediction is higher, making our proposed approach more efficient.

The tuning of the coherence factor in the cMSE loss function is of paramount importance for the proposed approach. The desired stability is application dependant and must, in the case of glucose prediction, be assessed by practitioners. In the future we plan on improving the loss function further by adding penalties directly tied to the CG-EGA (e.g., penalizing the model when the prediction is an EP). 

\subsubsection{Acknowledgments.}

We would like to thank the french diabetes health network Revesdiab and Dr. Sylvie JOANNIDIS for their help in building the IDIAB dataset used in this study.

\bibliographystyle{splncs04}
\bibliography{bibtex.bib}